\documentclass[aps,superscriptaddress,twocolumn,showpacs]{revtex4}
\usepackage{epsfig}
\usepackage{amsbsy}
\usepackage{amsmath}
\usepackage{graphicx}
\usepackage{latexsym}

\begin{document}

\title{Entanglement as a signature of quantum chaos}
\author{Xiaoguang Wang}
\affiliation{Department of Physics and Center for Nonlinear Studies, Hong Kong Baptist University, Hong Kong, China.}
\affiliation{Australian Centre of Excellence for Quantum Computer Technology, Department of Physics,    \\
Macquarie University, Sydney, New South Wales 2109, Australia.}

\author{Shohini Ghose}
\affiliation{Quantum Information Science Group, Department of
Physics and Astronomy, University of Calgary, Alberta T2N 1N4,
Canada.}

\author{Barry C Sanders}
\affiliation{Australian Centre of Excellence for Quantum Computer Technology, Department of Physics,    \\
Macquarie University, Sydney, New South Wales 2109, Australia.}
\affiliation{Quantum Information Science Group, Department of Physics and Astronomy, University of Calgary, Alberta T2N 1N4, Canada.}

\author{Bambi Hu}
\affiliation{Department of Physics and Center for Nonlinear Studies, Hong Kong Baptist University, Hong Kong, China.}
\affiliation{Department of Physics, University of Houston, Houston, Texas 577204-5005, USA.}

\date{\today}
\begin{abstract}
We explore the dynamics of entanglement in classically chaotic
systems by considering a multiqubit system that behaves
collectively as a spin system obeying the dynamics of the quantum
kicked top. In the classical limit, the kicked top exhibits both
regular and chaotic dynamics depending on the strength of the
chaoticity parameter~$\kappa$ in the Hamiltonian.  We show that
the entanglement of the multiqubit system, considered for both 
bipartite and pairwise entanglement, yields a signature of
quantum chaos. Whereas bipartite entanglement is enhanced in the
chaotic region, pairwise entanglement is suppressed. Furthermore,
we define a time-averaged entangling power and show that this
entangling power changes markedly as~$\kappa$ moves the system
from being predominantly regular to being predominantly chaotic,
thus sharply identifying the edge of chaos. When this entangling
power is averaged over initial states, it yields a signature of global
chaos. The qualitative behavior of this global entangling power is
similar to that of the classical Lyapunov exponent.
\end{abstract}
\pacs{05.45.Mt, 03.65.Ud, 03.67.-a}
\maketitle

\section{Introduction}

Quantization of classical chaotic systems has long been of
interest because of the challenges in identifying quantum
signatures of systems that, in a classical limit, exhibit chaotic
behavior. Various signatures have been identified, such as the
spectral properties of the generating
Hamiltonian~\cite{HaakeBook}, phase space scarring~\cite{Scar},
hypersensitivity to perturbation~\cite{Hyper}, and fidelity
decay~\cite{Fidelity}, which indicate an underlying chaotic
presence in the quantum dynamics. Here, we investigate the issue
of quantum signatures from the perspective of entanglement: as
entanglement is at the heart of quantum mechanics and a crucial
resource for quantum information processing~\cite{Nie00,Bennett},
the entanglement inherent in quantum chaotic systems could provide
a valuable approach to studying decoherence and quantum
chaos~\cite{Zurek93,Fur98,Mil99,Lak01,Ban02,Fuj03}. Furthermore,
quantum chaos could be seen as an engine for generating
entanglement; indeed quantum chaos could lurk in quantum
information processing~\cite{Bettelli03,SC2003,She00} as a deleterious or perhaps
even as an advantageous effect. We study entanglement by
considering a finite multipartite system, whose collective
dynamics obey chaotic Hamiltonian dynamics in the classical limit.

Previous studies of entanglement in chaotic systems
~\cite{Zurek93,Fur98,Mil99,Lak01,Ban02,Fuj03,Bettelli03, SC2003} have
explored bipartite entanglement in pure states, entanglement of
qubits in a multiqubit system and average entanglement or
entangling power. In this study we explore these different
types of entanglement in a single simple system, namely the quantum
kicked top~(QKT)~\cite{Haa86,Haa87,Fra85,Schack94,BarryChaos}.
This enables us to compare the degree to which these different
types of entanglement display signatures of chaos. 

An advantage of the dynamics of the QKT is that it obeys a spin algebra symmetry. This spin system can thus be regarded as a composite of distinct spin-half particles thereby admitting a multiqubit interpretation. Our system thus allows us to study and compare pairwise entanglement between two qubits as
well as bipartite entanglement between the two qubits and the rest
of the qubits. The Hilbert space for the QKT is finite and the
Poincar\'{e} section of the phase space compact, allowing analyses
of quantum and classical dynamics uncomplicated by truncation
issues. The QKT is well studied and understood thereby
simplifying the analysis of the role of entanglement in the
system. Finally, the QKT possesses a parity symmetry that allows
coherent quantum tunnelling to occur for states localized at
classical fixed points~\cite{BarryChaos}.

The Hamiltonian evolution may increase the entanglement of the
multipartite system, initiated in a collective spin coherent
state~\cite{SCSS}. For this analysis it suffices to employ two
entanglement measures. For bipartite entanglement, where the
multipartite system is divided into two subsystems, entropy of a
subsystem is used to quantify the degree of entanglement between
the two subsystems.  Pairwise entanglement, on the other hand,
considers the degree of entanglement between two qubits traced
over all remaining qubits and is quantified by the
concurrence~\cite{wootters1,wootters2}.

We present general results for both bipartite and pairwise entanglement
in the multipartite QKT and demonstrate that these entanglement
measures reveal strong signatures of the classical chaos features
corresponding to the onset of chaos and to the edge of
chaos~\cite{Edge}, which is the boundary between regular and
chaotic regimes in the classical phase space. We have studied the
behavior of the linear entropy and the concurrence for specific
initial states as well as the dynamics of these quantities when
averaged over all initial states.  We show that the linear entropy
increases more rapidly for an initial state centered in a chaotic
region of the classical phase space than one centered on a fixed
point. This agrees with the behavior of the linear entropy
observed in other chaotic systems, supporting the conjecture that
classical chaos can enhance bipartite
entanglement~\cite{Zurek93,Fur98}. Furthermore, we show that the
pairwise entanglement as measured by the concurrence also reveals
a dramatic change for a spin coherent state whose mean traverses
the edge of chaos on its transit through chaotic and regular
regions of the phase space. Contrary to the linear entropy, the
concurrence rapidly decreases for an initial state located in the
chaotic region.

While the linear entropy and concurrence can reveal the local
chaotic and regular structures in phase space,  the entangling
power, which is the averaged bipartite or pairwise entanglement, can
be used to identify the edge of chaos and quantify the onset of
global chaos, much like the classical Lyapunov exponent. We show
that the entangling power greatly increases as the chaoticity
parameter~$\kappa$ is increased, and the corresponding classical
kicked top makes the transition from predominantly regular to
predominantly chaotic behavior. In particular, the behavior of
the average linear entropy is qualitatively similar to that of the
classical Lyapunov exponent, thus revealing a signature of a
global feature of the classical chaos.

The paper is organized as follows. In Sec.~II, we introduce the
QKT and its classical dynamics, and introduce bipartite and
pairwise entanglement measures. In Sec.~III, we study in detail
dynamical evolutions of bipartite and pairwise entanglement, and
examine the edge of quantum chaos, the onset of quantum chaos via
the entangling power. We conclude in Sec.~IV.

\section{Background}

\subsection{Quantum kicked top}

The QKT is described by the Hamiltonian~\cite{Haa86,Haa87,Fra85}
\begin{equation}
H=\frac{\kappa}{2j\tau}J_z^2+pJ_y\sum_{n=-\infty }^{\infty }\delta
(t-n\tau ),  \label{top}
\end{equation}
where $J_{\alpha }~(\alpha \in \{x,y,z\})$ are spin operators and states are restricted to irrep $j$ for which $J^2=j(j+1)$. $\tau $ is the duration between periodic kicks, $p$ is the strength of each kick (which is manifested as a turn by angle $p$), and $\kappa$ is the strength of the twist. The Hamiltonian is an alternative sequence of twists ($J_z^2$ term) and turns ($J_y$ term). The QKT describes a spin system, which can be comprised of multiple systems of lower spins. For $\{\sigma_{i\alpha}\}$ the Pauli operators for the $i^{\text{th}}$ qubit, a collective spin operator, satisfying the usual su(2) algebra, is given by
\begin{equation}
J_{\alpha }=\sum_{i=1}^{N}\frac{\sigma _{i\alpha }}{2}.
\end{equation}
An example of using multiple qubits to simulate the QKT has been presented
for trapped ions~\cite{Milburn99}.

A standard dynamical description of the QKT is via the Floquet operator
\begin{equation}
F=\exp \left( \text{-i}\frac{\kappa }{2j\tau}J_{z}^{2}\right) \exp \left( \text{-i} p J_{y}\right) , \label{F}
\end{equation}
where the energy is rescaled so that $\tau=1$, and $p=\pi/2$ are henceforth assumed.
The orthogonal eigenstates of $F$, denoted by $\{|\Phi_m\rangle: -j\le m\le j\}$, which satisfy
\begin{equation}
F|\Phi_m\rangle=\exp(\text{i}\Phi_m)|\Phi_m\rangle,
\end{equation}
provide a convenient basis for stroboscopic evolution. An arbitrary state
$|\Psi (0)\rangle$ evolves to
\begin{equation}
|\Psi (n)\rangle =F^{n}|\Psi (0)\rangle=\sum_{m=-j}^{j}\Psi_m(0)\exp(\text{i}n\Phi_m)|\Phi_m\rangle.
\end{equation}
with $\Psi_m(0)=\langle\Phi_m|\Psi(0)\rangle$.

The QKT~\cite{Haa86,Haa87} is chaotic in the classical limit. For
integrable systems, it is well known from semi-classical theory
that the classical actions can be associated with corresponding
regular eigenstates of the quantum system with a well defined
quantum number. This correspondence breaks down in chaotic
systems~\cite{E1917}. In quasi-integrable systems with a mixed
classical phase space of regular and chaotic regions, some of the
eigenstates can still be associated with local actions in the
regular regions with corresponding discrete eigenenergies.
The remaining eigenstates result in an irregular energy spectrum
corresponding to the chaotic region~\cite{P1973}. We show here
that this underlying regular and chaotic energy spectrum of the
Floquet eigenstates of the QKT is reflected in the dynamics of the
entanglement, depending on whether the initial state is in the
regular or chaotic region of the classical phase space.

The classical limit of the QKT is obtained by expressing
$X=\langle J_x/j \rangle$ and similar for $Y$ and $Z$ and
factorizing all moments such as $\langle J_x J_y /j^2\rangle = XY$
to products of first-order moments. Then the classical equations
of motion, obtained from the Heisenberg operator equations of
motion and applying the factorization rule above, are given
by~\cite{Haa87}
\begin{eqnarray}
X^{\prime } &=&Z\cos (\kappa X)+Y\sin (\kappa X),  \notag \\
Y^{\prime } &=&-Z\sin (\kappa X)+Y\cos (\kappa X),  \label{map} \\
Z^{\prime } &=&-X  \notag
\end{eqnarray}%

The stroboscopic evolution described by Eq.(\ref{map}) can be
represented in a phase space given by a sphere $\mathcal{S}^2$ of
unit radius. The classical, normalized angular momentum variables
$(X,Y,Z)$ can be parametrized in polar coordinates as $
(X,Y,Z)=(\sin \theta \cos \phi ,\sin \theta \sin \phi ,\cos \theta
), $ where $\theta $ and $\phi $ are the polar and azimuthal
angles, respectively. Thus the mapping domain is essentially two-dimensional.

The stroboscopic dynamics of the classical map is shown in Fig.~1.
In the plot, we choose the chaoticity parameter $\kappa =3$ which yields a 
mixture of regular and chaotic areas of significant size.
Elliptic fixed points surrounded by the chaotic sea are evident.
Two such elliptic fixed points have coordinates $(\theta ,\phi
)=(2.25,-2.51)$ and $(\theta ,\phi )=(2.25,0.63).$ As we will see,
this phase space structure of the classical kicked top determines
behavior of quantum entanglement in the QKT.

\begin{figure}
\includegraphics[width=0.40\textwidth]{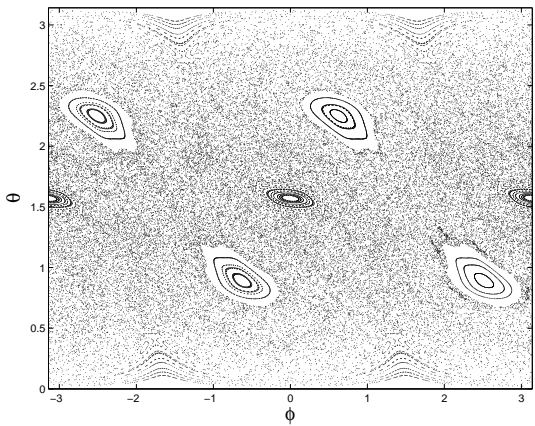}
\caption{The stroboscopic phase space dynamics of the classical kicked top
for $\kappa =3.$ Three hundred stroboscopic trajectories are
plotted, each for a duration of 300 kicks.}
\end{figure}

\subsection{Entanglement measures}

Pure-state bipartite entanglement has been calculated in previous
studies that explore connections between quantum entanglement and
underlying classical chaos~\cite{Fur98,Mil99,Lak01,Ban02,Fuj03}.
Quite recently, Bettelli and Shepelyansky~\cite{Bettelli03}
studied the behavior of the concurrence in a system exhibiting
quantum chaos, and found that the underlying classical chaos leads
to exponential decrease of the concurrence down to some residual
values. This result shows that the concurrence is very sensitive
to the onset of chaos, and understanding entanglement of these systems 
may help to control quantum chaos and suppress its
negative effects in quantum information processing.

We consider bipartite entanglement between a pair of
qubits selected from a symmetric multi-qubit state and the rest
of the system, as well as pairwise entanglement between the
two qubits. Once we obtain the two-qubit reduced density matrix,
the entanglement can be readily calculated. By expressing the
reduced density matrix $\rho _{12}$ for qubit 1 and 2 in terms of the expectation values of the collective operators, all elements of $\rho _{12}$ are
conveniently obtained~\cite{WangKlaus}.

Our system governed by the QKT Hamiltonian is a composite system,
and remains in a pure state at all times if we initially choose a
pure state. For pure states, bipartite entanglement is
well-defined and can be quantified by entropies of either
subsystem. For convenience, we adopt the linear entropy as the
entanglement measure, which is defined as
\begin{equation}
E=1-\text{Tr}_{1}(\rho _{1}^{2}),
\end{equation}
where $\rho_{1}$ is the reduced density matrix for the first
subsystem. The maximum linear entropy for a pure state of a
bipartite $d_1\times d_2$ system, is given by
$1-1/\min(d_{1},d_{2})$. While we may choose other entropies such
as the von Neumann entropy as our entanglement measure, the
qualitative results are in general independent of choice of
entropies for pure states. Moreover, the linear entropy and the
von Neumann entropy are two limiting cases of the R\'{e}nyi
entropy~\cite{Renyi},  they are thus interrelated and one can be used to estimate the other~\cite{Entropy1,Entropy2}.

Given our $N$-qubit system, we consider another type of
entanglement, the pairwise entanglement, i.e., the entanglement
between a pair of qubits. When $N\ge 3$, the pair of qubits can be
in a mixed state. Entanglement for a mixed state $\rho_{12}$ is
quantified by the entanglement of formation. Specifically, for a
pair of qubits, entanglement is equivalent to the non-positivity
of the partially transposed density matrix~\cite{peres}.
Alternatively, one can use the
concurrence~\cite{wootters1,wootters2} to quantify the pairwise
entanglement. The concurrence is defined as
\begin{equation}
\mathcal{C}=\max \left\{ 0,\lambda _{1}-\lambda _{2}-\lambda _{3}-\lambda
_{4}\right\} ,  \label{Cdef}
\end{equation}%
with the quantities $\lambda _{i}$ being the square roots of the eigenvalues
in descending order of the matrix product
$
\rho _{12}(\sigma _{1y}\otimes \sigma _{2y})\rho _{12}^{\ast
}(\sigma _{1y}\otimes \sigma _{2y}).
$
$\rho _{12}^{\ast }$ denotes the complex conjugate of $%
\rho _{12}$. The value of the concurrence ranges from zero for an unentangled state to unity for a maximally entangled state.

\section{Entanglement and quantum chaos}
We present here our studies of the entanglement dynamics of our
$N$-qubit system governed by the QKT~(\ref{top}), with the
relevant angular momentum operators $J_\alpha$ given by the
collective operators. If we choose the initial pure state to be
symmetric under exchange of any qubits, then the state vector at
any later time is also symmetric.
Thus, we can describe the state of the $N$%
-qubit system in terms of the orthonormal basis $\{|j,m\rangle;
(m=-j,-j+1,...,j)\}$ with $j=N/2$. The states $\{|j,m\rangle\} $ are the
usual symmetric Dicke states~\cite{Dic54}. State
$|j,-j\rangle $ is not entangled, whereas state $|j,-j+1\rangle $,
the so-called W state~\cite{Dur00,Wang1}, is pairwise entangled
with concurrence ${\cal C}=2/N$.

To connect the quantum and classical dynamics of the kicked top,
we choose the initial state to be the spin coherent state
(SCS) $\{|\theta,\phi\rangle 
=R(\theta ,\phi )|j,j\rangle; -\pi\le\phi\le\pi, 0\le \theta\le\pi\}$, 
with~\cite{SCSS}
\begin{equation}
R(\theta ,\phi ) =\exp \{\text{i}\theta \lbrack J_{x}\sin \phi -J_{y}\cos \phi
]\}.
\end{equation}%
Mean of $\mathbf{J}/j$ is 
\begin{equation}
\langle \theta ,\phi
|\mathbf{J}/j|\theta ,\phi \rangle =(\sin \theta \cos \phi ,\sin
\theta \sin \phi ,\cos \theta ). 
\end{equation}
The initial SCS can be
rewritten as a multi-qubit product state, and thus exhibits no
entanglement (zero linear entropy and concurrence).

\subsection{Dynamics of entanglement}

We start by exploring the dynamics of entanglement for initial
states with a mean value in four different regions of the phase space,
specifically a fixed point, an integrable (or KAM) region, a
chaotic region, and the border between the integrable and the
chaotic region known as `the edge of chaos'~\cite{Edge}. We are also
interested in the behavior as the chaoticity parameter $\kappa$ is
varied. We start with the choice $\kappa=3$, which exhibits large
integrable and large chaotic regions, and we select four states
localized in the four regions mentioned earlier.

For convenience we fix $\theta=2.25$ and vary $\phi$: this `line
of latitude' on $\mathcal{S}^2$ includes all four regions we are
exploring. An elliptic fixed point arises at $\phi=0.63$, a point
in the regular region occurs at $\phi=0.90$, one edge of chaos can
be seen at $\phi=1.05$, and a point well in the chaotic sea is
located at $\phi=2.00$. The states with means at each of these
$(\theta,\phi)$ points in the phase space are chosen to be SCSs.
These states are minimum uncertainty states and are well localized
around the four chosen points in phase space.

\begin{figure}
\includegraphics[width=0.40\textwidth]{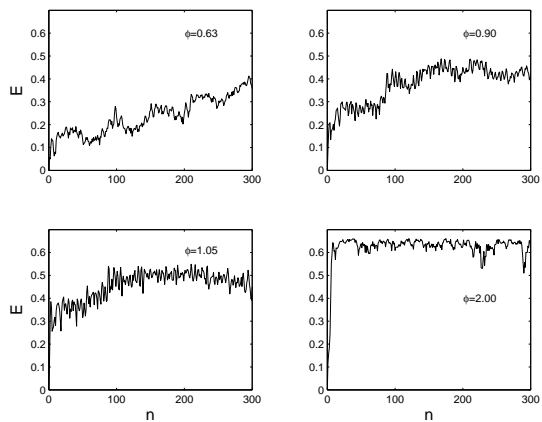}
\caption{ Dynamical evolution of the linear entropy for initial SCS $|\theta=2.25,\phi$ for $\kappa =3$ and $N=50$.}
\end{figure}

We numerically compute bipartite entanglement between two qubits
and the other $N-2$ qubits, and display the numierical results in Fig.~2 as linear entropy versus time increasing parameter $n$. We observe that the entanglement is enhanced for the initial state centered in the
chaotic region after a short time. Initially, the linear entropy is zero, and as the dynamics evolve, the entropy increases slowly when the wave packet
is centered in the regular region, whereas it exhibits a rapid
rise when centered in the chaotic region. The curve with $\phi
=1.05$ displays the intermediate behavior. Furthermore, the
entanglement for the state initially with mean at a fixed point
displays a periodic modulation that is absent in the evolution of
the entanglement for the state initiated in the chaotic region.
This periodic modulation is an indicator of the underlying regular
classical dynamics and  corresponding regular energy spectrum of
the Floquet eigenstates~\cite{P1973} .

\begin{figure}
\includegraphics[width=0.40\textwidth]{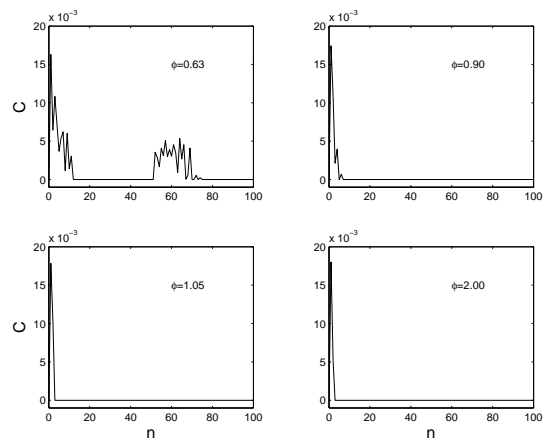}
\caption{ Dynamical evolution of the concurrence. The parameters are the same
as those of Fig. 2.}
\end{figure}
\begin{figure}
\includegraphics[width=0.40\textwidth]{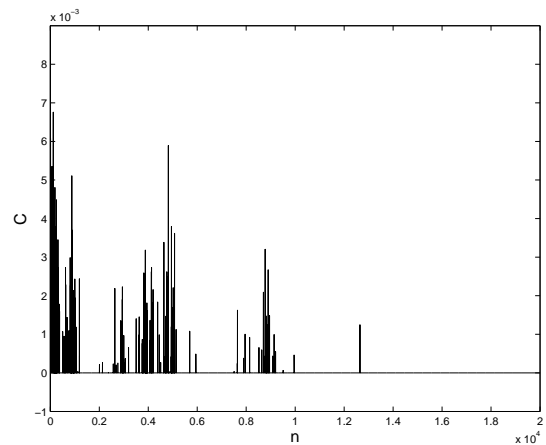}
\caption{Long time behavior of the concurrence for the regular case ($\theta=2.25, \phi=0.63$).  The other parameters are the same
as those of Fig. 2.
}
\end{figure}

Figure 3 shows the dynamical behavior of the concurrence. Just as
in the linear entropy case, we see a rapid change in the
concurrence for a state initially centered in the chaotic sea.
However, our numerical results suggest that the initial state
centered in the fixed point will lead to large pairwise
entanglement production, which is opposite to the case of the
bipartite pure-state entanglement production, in which the
classical chaos enhances the production of bipartite entanglement.
Of particular interest are the collapses and revivals in the evolution of the concurrence for a state centered on
the fixed point. Figure 3 (left-up one) shows a revival at $n=52$. Additional revivals occur at $n=113$, $n=183$ and so on. For the chaotic case, we cannot observe the
revival phenomenon. This quasiperiodic behavior in the regular
region indicates that the SCS in the regular region has a
finite support over the basis set of regular Floquet eigenstates.

In Fig.~4 we plot the long-time behaviors of the concurrence for
the regular case, and observe multiple collapse and revivals. At
long times the revivals become sparse, and finally the concurrence
reduces to zero. We see that the concurrence is very sensitive to
quantum chaos, which is consistent with the observations of
Bettelli and Shepelyansky~\cite{Bettelli03} in their studies of
concurrence between qubits during the operation of an efficient
multiqubit quantum algorithm. However, unlike their system in
which the concurrence reached a finite residual value, in our QKT
model, the concurrence disappears at very long times; the difference arises because we assumes symmetrised multi-qubit states, whereas Bettelli and
Shepelyansky allow this symmetry to be broken.

\subsection{Edge of quantum chaos}

The edge of chaos~\cite{Edge} is an important issue in the study of
quantum chaos. In classical chaos, the edge of chaos is a fractal
boundary separating the regular and chaotic regions. However, this
fine-grained fractal structure does not translate well into the
quantum domain. Recently, it was found that the edge of chaos is
characterized by a power law decrease in the overlap between a
state evolved under chaotic dynamics and the same state evolved
under a slightly perturbed dynamics~\cite{Edge}. Here, we study
the edge of quantum chaos from the perspective of entangling
powers, which are defined to be either the maximal or the mean entanglement
that the evolution operator can generate over all initial
states~\cite{P,Paolo}. Alternatively, given a fixed initial state,
we may ask what is the maximal and the mean entanglement that the
operator can generate over all time. In general state-averaging and time-averaging are inequivalent and so the two methods yield different results.

In strongly chaotic systems,
the two definitions converge due to nearly ergodic dynamics. In this study, we explore both the entanglement averaged over all time as well as the entanglement averaged over initial states. We begin our analysis with the average over time. In practice, for numerical purposes we consider a finite time domain. We define time-averaged entanglement power as the
average linear entropy or average concurrence over a time interval $T$ (which should be much longer than other time scales) as follows:
\begin{equation}
E_{T}=\frac{1}{T}\int_0^T \text{d}t~E(t),\, {\cal C}_{T}=\frac{1}{T}\int_0^T \text{d}t~{\cal C}(t).
\end{equation}
For local unitary operations, the above quantities are necessarily
zero.
\begin{figure}
\includegraphics[width=0.40\textwidth]{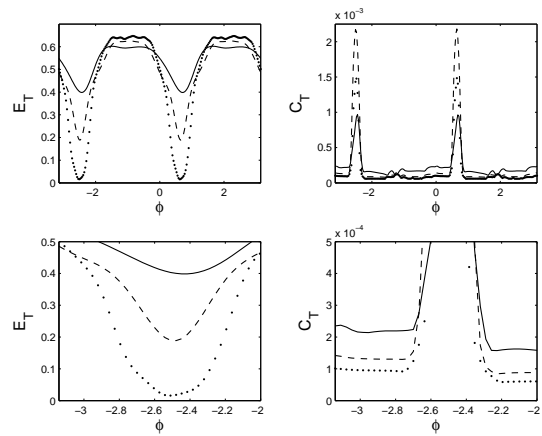}
\caption{Mean linear entropy and mean concurrence against
azimuthal angle $\phi$. The mean linear entropy is plotted for
$N=20$ (solid line), $N=50$ (dashed line), and $N=100$ (dotted
line); the mean concurrence is plotted for $N=100$ (solid line),
$N=200$ (dashed line), and $N=300$ (dotted line). Parameters
$\theta$ and $\kappa$ are set to $\theta =2.25$ and $\kappa=3$. 
The time average is over 200 steps, and the subplots below are 
partly enlarged versions of above ones.}
\end{figure}

We fix the polar angle $\theta =2.25$ of the SCS as before, and
vary the azimuthal angle $\phi $ from $-\pi $ to $\pi$.  The
center of the SCS wave packet thus commences in the chaotic region and passes through two
regular islands. Figure 5 displays the time-averaged mean linear entropy $E_T$and mean concurrence $C_T$ as a function of
the azimuthal angle $\phi$. When the azimuthal angle goes from $-\pi $ to
the first regular region, the linear entropy decreases until it
reaches a minimum which approximately corresponds to  the fixed
point $(\theta ,\phi )=(2.25,-2.51)$. Subsequently the mean
entropy increases to a flat larger area corresponding to the
chaotic region.

In contrast to the behavior of the mean linear entropy, the mean
concurrence reaches a maximum approximately at the fixed point.
When $\phi $ increases from $-\pi$, the mean concurrence first
decreases slowly,  and then exhibits an abrupt increase to a
maximal value. Thos turning point sharply displays the edge of
chaos. Another turning point is obvious from the figure. 

We also observe that the larger the number of qubits, the
wider the regular region. In the large $N$ limit, the regular
region will coincide with that of corresponding classical chaos of
Fig.~1. 

We further calculate the mean linear entropy and the mean
concurrence as a function of $\theta$ and $\phi$. The contour
plots are shown in Figs.~4 and 5. Comparing Figs.~1 and 4, we
observe that these two figures closely match each other.
Specifically, the four islands of Fig.~4 are evident, reflecting
the four stable islands in the classical phase space. Comparing
Figs.~1 and 5, the four stable islands of Fig.~1 closely match
those of Fig.~5. Thus, we have a good classical-quantum
correspondence.

\begin{figure}
\includegraphics[width=0.40\textwidth]{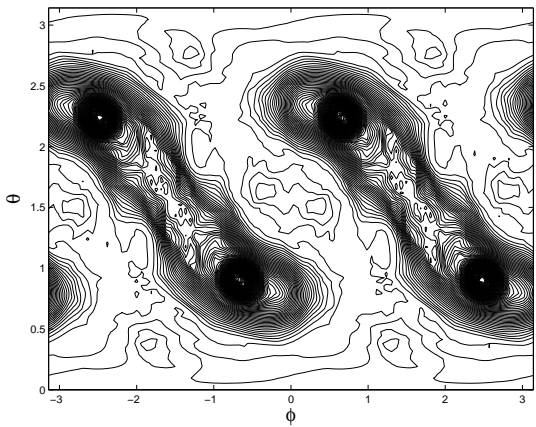}
\caption{Contour plot of mean linear entropy against $\phi$ and
$\theta$. The mean is over 200 kicks and other parameters are the
same as those in Fig.~2.}
\end{figure}

\begin{figure}
\includegraphics[width=0.40\textwidth]{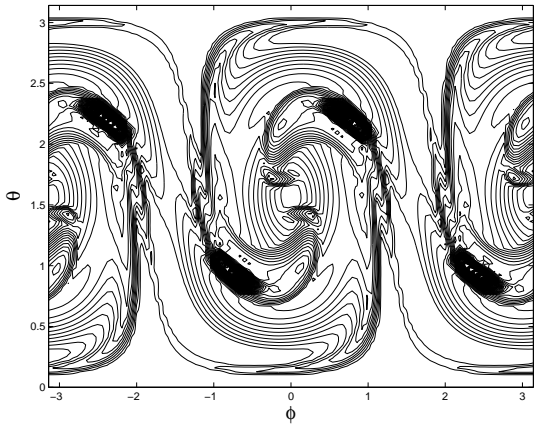}
\caption{Contour plot of mean concurrence against $\phi$ and $\theta$.
The average is over 200 kicks and other parameters are the same as those in Fig.~2.}
\end{figure}

\subsection{Onset of Chaos}
\begin{figure}
\includegraphics[width=0.40\textwidth]{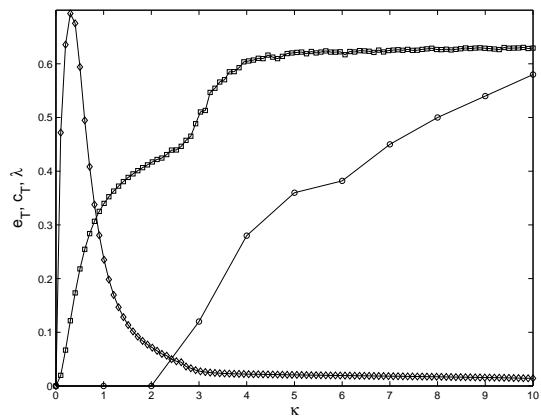}
\caption{Entangling power $e_T$ (square line), $c_T$ (diamond
line), and the Lyapunov exponent $\lambda$ (circle line)agaist
$\kappa$. The plot of Lyapunov exponent corresponds to fig.~1 of
Ref.~\cite{LY}. The parameter $N=36$ and time average is over 50
kicks.}
\end{figure}
In the previous section, the time-averaged entanglement revealed clearly whether the initial state was in the regular or chaotic region. Here, we are concerned with global properties of a chaotic system.
For classical systems, the onset of global chaos can be quantified by
calculating the global Lyapunov exponent. Here we
define the following entangling power to quantify the onset of global
quantum chaos:
\begin{align}
e_{T}=\frac{1}{T}\int \text{d}\mu\int_0^T \text{d}t~E(t,\theta,\phi),\nonumber\\
{c}_{T}=\frac{1}{T}\int \text{d}\mu \int_0^T \text{d}t~{\cal C}(t,\theta,\phi),
\end{align}
where $\text{d}\mu=\text{d}\mu(\theta,\phi)$ is the Haar measure. Like the
global classical Lyapunov exponent, this state-averaged entangling power characterizes global properties of the QKT. The entangling power characterizes the entangling capability of the $\kappa$-dependent Floquet operator.

Figure 8 shows the entangling powers and the Lyapunov exponent
versus parameter $\kappa$. We observe that when $\kappa=\kappa_0\approx
2.4$, the entangling power $e_T$ exhibits a rapid increase and
saturates beyond $\kappa=\kappa_1\approx 5$. The rapid increase of the entangling power signifies
the onset of quantum chaos, and the saturation  implies that
global chaos has occurred. Between $\kappa_0$ and $\kappa_1$, when
there are still regions of regular islands in the chaotic phase
space, a mixture of regular and chaotic behavior is expected. In
contract to $e_T$, $c_T$ becomes very small for $\kappa
> \kappa_0$, which is also an indicator of onset of quantum chaos.
Note that $c_T$ has a peak which results from the competition
between the entangling power of the QKT Floquet operator and the
inherent chaos. On one hand, increasing $\kappa$ will enhance the
entangling power, and on the other hand, the inherent quantum
chaos suppresses the pairwise entangling power, thus leading to
the peak. For $e_T$, increase of $\kappa$ and the quantum chaos
both enhance the linear entropy, and thus no competition exists
and no peak appears.

\section{Conclusions}
We have investigated a multi-qubit system whose collective Hamiltonian
dynamics are chaotic in the classical limit. We studied the particular
example of the quantum kicked top, which is a well-studied example of
quantum chaos with the advantages of having a finite-dimensional Hilbert
space (thereby obviating the need for truncation that
arises in infinite-dimensional Hilbert spaces) and of involving only spin
operators of no more than quadratic order.

Although the collective dynamics are well understood, the underlying
entanglement of the qubits that collectively make up the quantum kicked
top is only just beginning to be understood. Here we have developed
methods for studying the quantum kicked top, and these methods are
applicable to more general systems. We have identified bipartite and
pairwise entanglement as two quite distinct measures to determine the
entanglement in the system, and we have related the dynamics of these
measures of entanglement to chaotic features of the quantum kicked top in
the classical limit; as examples, we have connected the features of the
entanglement evolution to local properties such as whether the state is
supported predominantly in the regular or chaotic region and also to
global properties such as showing that entangling power averaged over
states grows similarly to the global Lyapunov exponent growth for the classical
chaotic system. 

We have assumed symmetric multi-qubit states throughout, and the
entanglement properties studied here reflect this assumption. If the
symmetrization condition is broken, different dynamics can be expected.
For example, Bettelli and Shepelyanski~\cite{Bettelli03} show a concurrence that reaches a residual steady-state value. They explain this non-zero residue as being a result of symmetry breaking in their system. In contrast, our system exhibits a decay of concurrence to zero. The assumption of symmetric states implies indistinguishability of the qubits.   Thus even though entanglement may exist in the system, it may not be accessible as a useful tool for quantum information processing due to inherent inability to distinguish between the qubits. An alternative measure of entanglement could be an operational measure that takes into account physical restrictions on accessibility of the entanglement due to symmetries of the system~\cite{WV2003}.  

In summary, our work highlights the connection between entanglement of a multi-qubit state whose collective dynamics is chaotic in the classical limit and
introduces valuable methods and measures for studying this entanglement. It would be worthwhile to investigate other quantum chaotic systems using concepts of time-averaging and global entangling power. Furthermore, it would interesting to define and compare systems in which the symmetrization results hold to systems where this symmetrization condition is broken. 

\acknowledgments We acknowledge valuable discussions with L.T.
Stephenson, X.W. Hou, H.B. Li, B.S. Xie, L. Yang, and H. Zhang.
This project has been supported by an Australian Research Council
Large Grant, the Hong Kong Research Grants Council (RGC), a Hong
Kong Baptist University Faculty Research Grant (FRG), and
Alberta's informatics Circle of Research Excellence (iCORE).

\end{document}